\documentclass[superscriptaddress,preprintnumbers,byrevtex,floatfix,twocolumn]{revtex4}
\usepackage[T1]{fontenc}
\usepackage[latin9]{inputenc}
\usepackage{amstext}
\usepackage{graphicx}
\usepackage{amssymb}
\usepackage{upgreek}

\begin{document}
\title{On the application of radio frequency voltages to ion traps via helical resonators}
\author{J.D. Siverns, L.R. Simkins, S. Weidt, W.K. Hensinger\footnote{Corresponding author. Email: W.K.Hensinger@sussex.ac.uk
\vspace{6pt}}}
\affiliation{Department of Physics and Astronomy, University of Sussex, Brighton}

\begin{abstract}
Ions confined using a Paul trap require a stable, high voltage and low noise radio frequency (RF) potential. We present a guide for the design and construction of a helical coil resonator for a desired frequency that maximises the quality factor for a set of experimental constraints. We provide an in-depth analysis of the system formed from a shielded helical coil and an ion trap by treating the system as a lumped element model. This allows us to predict the resonant frequency and quality factor in terms of the physical parameters of the resonator and the properties of the ion trap. We also compare theoretical predictions with experimental data for different resonators, and predict the voltage applied to the ion trap as a function of the Q-factor, input power and the properties of the resonant circuit.
\end{abstract}
\maketitle

\section{Introduction}

Trapped ions are a powerful tool which have many applications such as mass spectrometry \cite{Paul} and frequency standards \cite{Bollinger,Fisk}, whilst also being recognised as a leading contender for the practical implementation of quantum information processing \cite{Kieplinski,Cirac,Haffner} and quantum simulations \cite{Clark,Ivanov,Porras,Schmitz}. To trap ions within a Paul trap a high radio frequency voltage is applied to electrodes in order to provide the required electric potentials. A helical resonator allows impedance matching between a radio frequency source and an ion trap enabling high voltages while reducing the noise injected into the system. These properties make the resonator an important device not only in ion trapping but also in a wide range of physical sciences including ultra high frequency (UHF) mobile communication systems \cite{Yu}, spin resonance spectroscopy \cite{Collingwood} and measuring the dielectric properties of materials \cite{Meyer}. 

An empirical study of shielded helical coil resonators was performed by Macalpine and Schildknecht \cite{Macalpine} who considered isolated operation at a self resonant frequency due the the coil inductance and shield capacitance. In contrast we consider a shielded helical coil connected to an ion trap where the resonant frequency will be determined by the whole system. Macalpine and Schildknecht \cite{Macalpine} showed that, when tuning a resonator with an external capacitance, the $Q$ factor would vary with the tuned resonant frequency, however, they did not predict this resonant frequency or the effect of a lossy (resistive) capacitance on the $Q$ factor. Due to these limitations, when connecting an ion trap to a helical resonator the predications of Macalpine and Schildknecht \cite{Macalpine} for the resonant frequency and $Q$ factor can deviate by orders of magnitude from those observed. In this paper we will provide reliable predictions for the resonant frequency and $Q$ factor for a shielded helical coil connected to an ion trap impedance. We also provide a design guide which allows a helical resonator to be constructed with the highest possible $Q$ factor for the constraints of a particular experiment. We also discuss the process of impedance matching with a helical resonator and predict the voltage applied to the ion trap as a function of the Q-factor, input power and the properties of the resonant circuit.

\section{Trapping Charged Particles}

\label{sec:trapping} 
\begin{figure}\label{trap}
\centering \includegraphics[width=1\columnwidth]{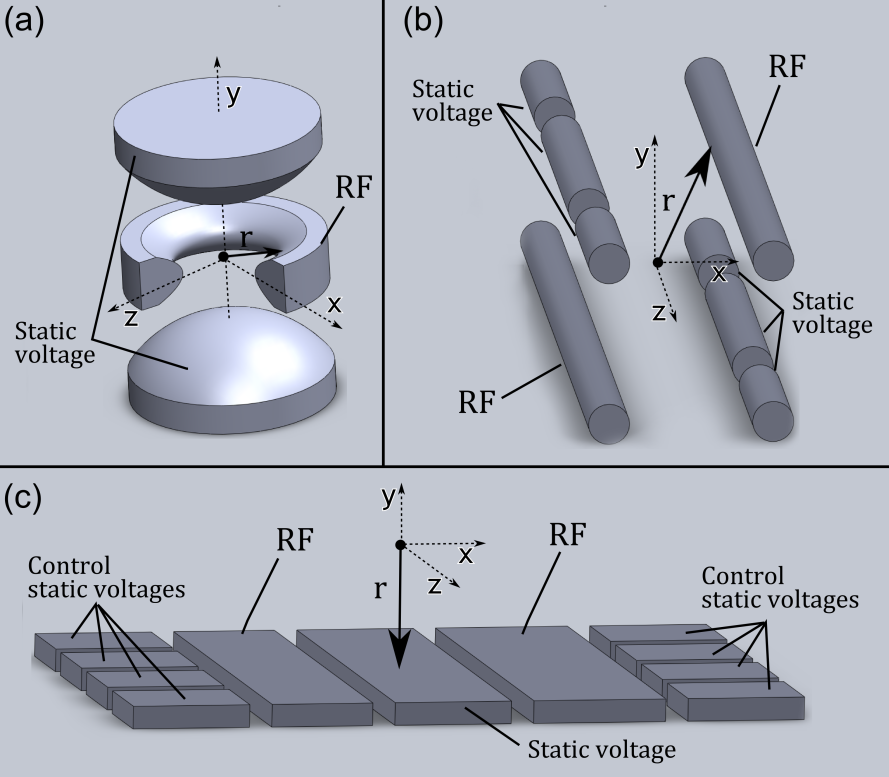}
\caption{(a) Diagram showing a quadrupole Paul trap. The ponderomotive potential due to the RF and ground electrodes provide confinement in all three dimensions. All electrodes are hyperbolically shaped. (b) A two-layer linear Paul trap with two rod RF electrodes and two segmented RF ground electrodes providing transverse confinement with static voltages present to provide confinement in the axial dimension. (c) Diagram showing a five-wire surface Paul trap. Here the ion is confined in a potential created by the surface electrodes at the mid point between the two RF strips. Control static voltage electrodes are present to enable confinement along the axial direction of the trap. This enables ions to be transported along the length of the trap \cite{transport}.}
\end{figure}

Paul traps \cite{Paul}, which are a prime candidate for quantum information processing \cite{Kieplinski,Cirac,Haffner}, trap ions using a radio frequency (RF) voltage applied to some of the trap electrodes in order to obtain a suitable ponderomotive pseudo-potential. Figure \ref{trap} shows examples of different types of Paul trap designs. The potential used to trap an ion of mass, $m$, and charge, $e$, is given by \cite{ghosh,Madsen}
\begin{equation}\label{pond}
\psi=\frac{e^{2}V_{0}^{2}\eta^{2}}{4mr^{4}\Omega_{rf}^{2}}\left(x^{2}+y^{2}\right),
\end{equation}

\noindent
where $\Omega_{rf}$ is $2\pi$ times the RF frequency in Hertz (throughout the text, when analysing the overall resonant circuit we will refer to $\Omega_{rf}$ as $\omega$ or $\omega_0$ when the circuit is on resonance), $r$ is the distance from the centre of the trap to the nearest electrode and $V_{0}$ is the amplitude of the RF voltage applied to the trap. $\eta$ is a geometric efficiency factor \cite{Madsen} which is equal to one for a perfectly hyperbolic geometry (similar to the trap shown in figure \ref{trap}(a)) and less than one as the geometry strays from this perfect form.

The secular frequency of an ion trapped inside the ponderomotive pseudopotential, given in equation \ref{pond}, is then given by \cite{ghosh,Madsen}

\begin{equation}\label{secular}
\omega_{s}=\frac{eV_{0}\eta}{\sqrt{2}mr^{2}\Omega_{rf}}.
\end{equation}

Linear ion traps (as seen in figure \ref{trap}(b) and (c)) possess a linear node in the produced ponderomotive potential. To provide ion confinement along this node additional static voltages are applied to certain electrodes. The equations of motion for an ion inside this potential are given by the Mathieu equations \cite{Paul}, which feature stability parameters. Stable motion of an ion in the trapping field will only occur over certain ranges of these parameters and is dependent on the initial position and momentum of the ion. It can be seen from equations \ref{pond} and \ref{secular} that a high voltage will lead to a deep trapping potential and a high secular frequency. A deep potential provides long ion lifetimes and large secular frequencies typically allow for faster ion transportation \cite{transport}, more efficient ground state cooling and shorter quantum gate times.

The application of high voltages must, however, be limited to avoid both electrical breakdown or experimentally intrusive temperatures due to the power dissipated in the ion trap. It is important to know the voltage being applied to the ion trap for a given input power. A combined ion trap-resonator system can be represented as a series LCR circuit with resonant frequency $\omega_{0}=\frac{1}{\sqrt{LC}}$ and $Q=\frac{1}{R}\sqrt{\frac{L}{C}}$. The voltage over the ion trap will be approximately equal to the voltage over the capacitor when the capacitance of the ion trap dominates the overall capacitance of the circuit. At resonance the peak voltage over the capacitor is equal to the peak voltage over the inductor. The instantaneous voltage of the inductor is

\begin{equation}\label{eq:voft}
V(t)=L\frac{dI_{peak}\sin(\omega_{0}t)}{dt}=L\omega_{0}I_{peak}\cos(\omega_{0}t),
\end{equation}
where $I_{peak}$ is the peak current and $L$ is the coil inductance.
The peak voltage over the inductor occurs when $\cos(\omega_{0}t)=1$, hence the peak voltage over the ion trap can be approximated as

\begin{equation}\label{eq:vpeak}
V_{peak}\approx LI_{peak}\omega_{0}.
\end{equation}

\noindent
Power is only dissipated in the system through the resistance $R$, thus the power dissipated is

\begin{equation}\label{eq:pd}
P_{d}=RI_{rms}^{2}=\frac{1}{2}RI_{peak}^{2},
\end{equation}

\noindent
where $I_{rms}=\frac{1}{\sqrt{2}}I_{peak}$ is the root-mean-square current. Using these equations and $Q=\frac{1}{R}\sqrt{\frac{L}{C}}$, we find that

\begin{equation}\label{outV2}
V_{peak}\approx\kappa\sqrt{2PQ},
\end{equation}

\noindent
where,

\begin{equation}\label{kappa}
\kappa=\left(\frac{L}{C}\right)^{\frac{1}{4}},
\end{equation}

\noindent
and $V_{rms}=V_{peak}/\sqrt{2}\approx\kappa\sqrt{PQ}$.

This shows the output voltage of a resonating system can be predicted given the input power, $P$, the capacitance, $C$, inductance, $L$, and quality factor $Q$ of the system. Applying RF voltages via a high $Q$ factor resonator reduces the power in unwanted frequencies being applied, reducing their contribution to motional heating of ions \cite{Turchette} and also provides higher voltages per input power, resulting in deeper trapping potentials and higher secular frequencies.

The impedance of the ion trap and connections are typically large enough to contribute to the response of an LCR resonator, and thus must be considered when designing a resonator to operate at a given frequency. Considering $\omega_{0}=\frac{1}{\sqrt{LC}}$ and $Q=\frac{1}{R}\sqrt{\frac{L}{C}}$, in order to maximise the $Q$ factor for a fixed frequency $\omega_{0}$ we must minimise $C$ while maximising $L$. The use of a helical coil allows for an inductor to be made with a low self capacitance and resistance, enabling the resistance and capacitance of the ion trap and connections to dominate the $R$ and $C$ of the LCR resonator and thus maximising the $Q$ factor.

\section{Circuit model}
\subsection{Impedance matching via inductive coupling}\label{sec:couple}

RF voltages can be applied by direct connection from the ion trap to an RF amplifier, however this can cause multiple issues. A mismatch of impedance between the amplifier and the ion trap will cause the RF signal to be reflected from the ion trap, resulting in power dissipated over the output impedance of the amplifier. This will require an RF amplifier with a much greater power handling than for a matched system. The amplifier will also inject noise into the ion trap which can cause heating of the ion \cite{Turchette}. Passing the output of the amplifier through a resonator will filter this noise, reducing the contribution to ion heating. In order to maximise the filtering of this noise the resonator must have a high $Q$ factor, and hence a narrow bandwidth. Direct connection of a resonator to the amplifier will reduce the resonator's $Q$ factor due to the damping effect of the finite output impedance of the amplifier. The RF amplifier can, however, be connected to the resonator through a capacitive or inductive coupling, which decouples the resonator from the resistive output impedance of the amplifier, allowing for a resonator with a high $Q$ factor. This technique also allows impedance matching of the ion trap and RF amplifier by altering the coupling, thus reducing the reflected power, and hence the required power for a given voltage.

\begin{figure}
\centering \includegraphics[width=1\columnwidth]{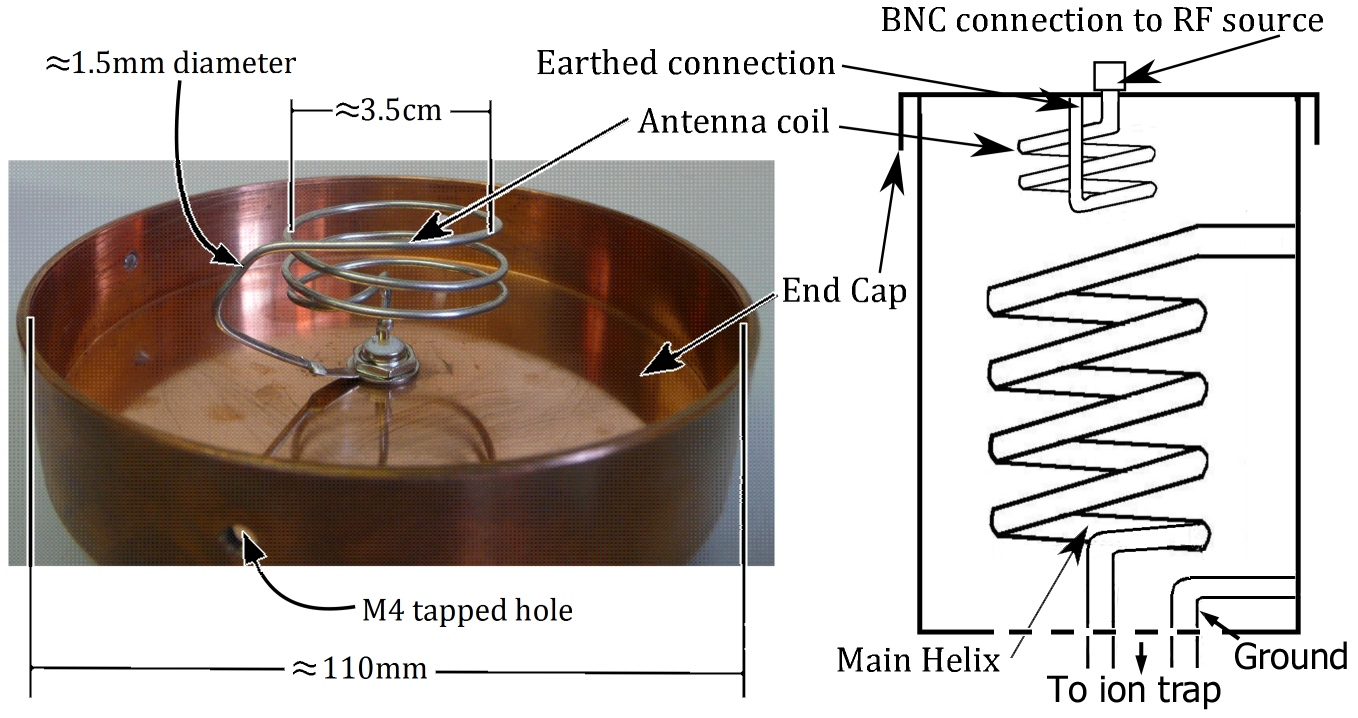}
\caption{A resonator end cap showing the antenna coil together with a diagram showing its location in a fully constructed helical resonator.}
\label{antenna}
\end{figure}

For inductive coupling an antenna coil is attached to an end cap and placed directly and centrally above the main helical coil as shown in figure \ref{antenna}. By altering the physical properties of this coil, impedance matching between the resonator and the radio frequency source can be achieved.

\begin{figure}
\centering \includegraphics[width=1\columnwidth]{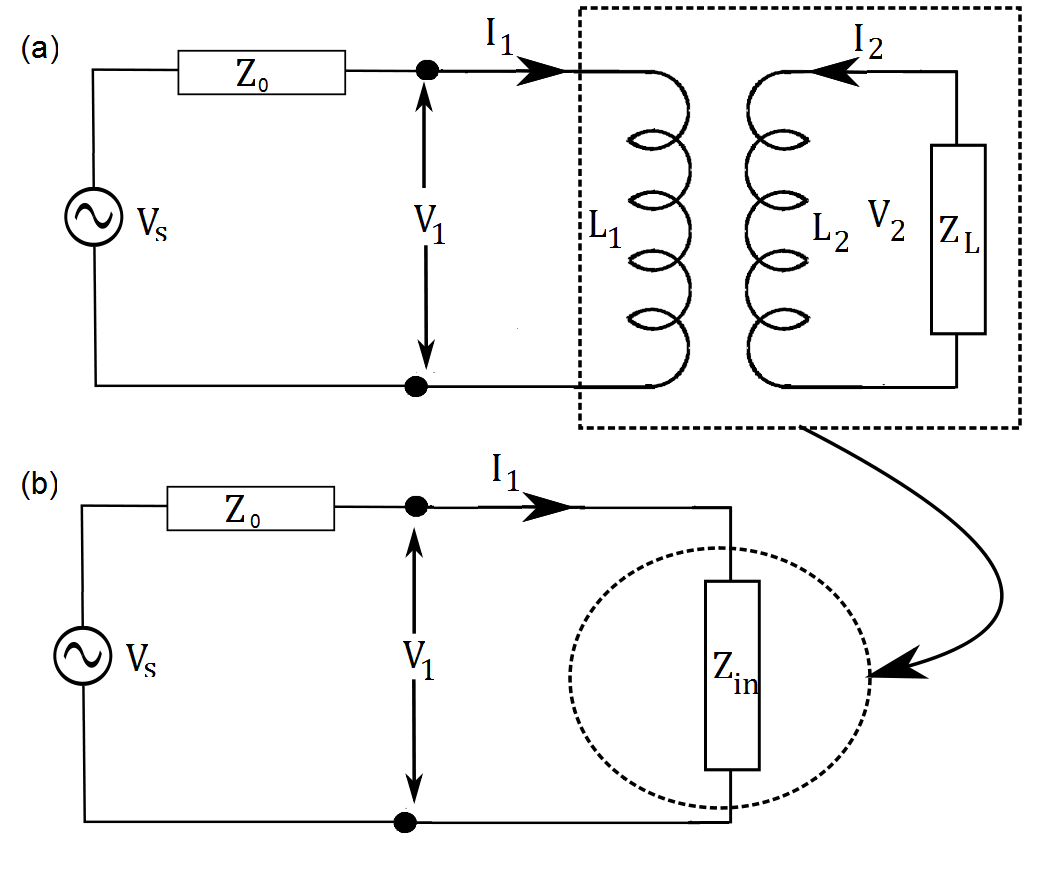}
\caption{(a) Circuit of a resonator with attached load (inside dashed box) and signal generator (outside dashed box) represented as two physically separate circuit loops coupled together by the antenna coil inductor, $L_{1}$, and the main coil inductor, $L_{2}$. The impedance of the ion trap is shown as a load impedance, $Z_{L}$. The source voltage and impedance (outside dashed box) are shown as $V_{s}$ and $Z_{0}$ respectively. (b) The circuit from (a) is represented as the Thévenin equivalent impedance, $Z_{in}$, along with the source voltage and impedance.}
\label{inductors}
\end{figure}

To understand how altering the physical properties of the antenna coil allows impedance matching, the resonator is represented by two electrically isolated circuit loops as shown in figure \ref{inductors}. Here the inductor, $L_{1}$, represents the antenna coil and the inductor, $L_{2}$, represents the main coil. The voltage source $V_{s}$ and impedance $Z_{0}$ represent the output of an RF amplifier. The two coils are placed in close proximity to each other creating a coupling between the two circuit loops due to the mutual inductance.

The voltage in each circuit loop is given by

\begin{equation}\label{faraday}
V_{1/2}=i\omega L_{1/2}I_{1/2}+i\omega MI_{2/1},
\end{equation}

\noindent
where $M=k\sqrt{L_{1}L_{2}}$ is the mutual inductance of the two coils and $k$ is the coupling. The equivalent circuit is given in figure \ref{inductors}(b) where the impedance $Z_{in}$ describes the overall impedance of the resonator (and ion trap), which can be adjusted by altering the physical parameters of the antenna coil enabling an impedance match to the RF amplifier ($V_{s}$ and $Z_{0}$). The overall impedance, $Z_{in}$, of the two circuits, as shown in figure \ref{inductors}, is then

\begin{equation}\label{zin1}
Z_{in}=V_{1}/I_{1}=i\omega L_{1}+i\omega M\frac{I_{2}}{I_{1}}.
\end{equation}

Using

\begin{equation}\label{v2}
V_{2}=-Z_{L}I_{2}
\end{equation}

\noindent
and equation \ref{faraday} we obtain

\begin{equation}\label{det2}
(Z_{L}+i\omega L_{2})I_{2}+i\omega MI_{1}=0.
\end{equation}

Rearranging equation \ref{det2} for $I_{2}/I_{1}$ gives

\begin{equation}\label{eq:11}
\frac{I_{2}}{I_{1}}=\frac{-i\omega M}{Z_{L}+i\omega L_{2}}.
\end{equation}

Inserting equation \ref{eq:11} into equation \ref{zin1} we can describe the overall impedance as

\begin{equation}\label{zin2}
Z_{in}=i\omega L_{1}+\frac{\omega^{2}M^{2}}{i\omega L_{2}+Z_{L}}.
\end{equation}

We can approximate the antenna coil's inductance as $L_{1}=\frac{\mu_{0}NA}{\tau}$, where $\tau$, $N$ and $A$ are the winding pitch, number of turns and cross sectional area of the coil respectively and $\mu_{0}$ is the permittivity of vacuum, giving

\begin{equation}\label{zin_antenna}
Z_{in}=\frac{\mu_{0}NA}{\tau}\left(i\omega+\frac{k^{2}L_{2}\omega^{2}}{i\omega L_{2}+Z_{L}}\right).
\end{equation}

\begin{figure}
\centering \includegraphics[width=1\columnwidth]{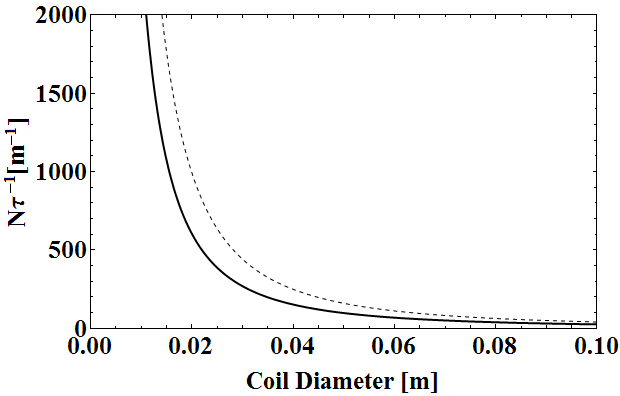}
\caption{The ratio of number of antenna turns $N$ to winding pitch $\tau$, required to impedance match the load to a 50 Ohm source is plotted as a function of the antenna coil diameter. This is shown for the case when the combined resonator-ion trap load is high (solid black line, a resistance of 15 Ohm and a capacitance of 100 pF), and for the case when the load is small (dashed black line, a resistance of 0.2 Ohm and a capacitance of 1 pF). In both cases a resonant frequency of $\omega=2\pi\times20$ MHz is used.}
\label{antratio}
\end{figure}

\begin{figure}
\centering \includegraphics[width=1\columnwidth]{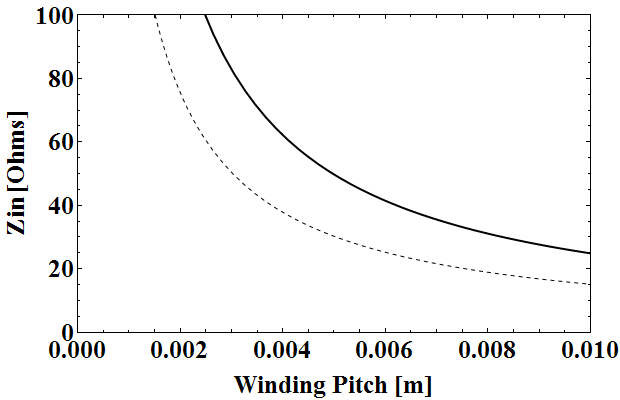}
\caption{The resonator input impedance, $Z_{in}$, is shown as a function of the winding pitch, $\tau$, of the antenna coil. This is shown for the case of a high combined resonator-ion trap load ($Z_{L}$ given by a resistance of 15 Ohm  and a capacitance of 100 pF) shown by the solid black curve and for the case of a small load ($Z_{L}$ given by a resistance of 0.2 Ohm  and a capacitance of 1 pF) shown by the dashed black curve. In both cases a resonant frequency of $\omega=2\pi\times20$ MHz is used and the number of turns and coil diameter are kept constant at 3 and 3 cm respectively.}
\label{antpitch}
\end{figure}

This shows that the input impedance of the resonator can be altered by simply adjusting the physical parameters of the antenna coil and, thus, match it to that of the voltage source. To illustrate how the physical parameters have to be altered to achieve a matching, equation \ref{zin_antenna} has been plotted for two common cases. The first case (solid black line) is when the combined resonator - ion trap load is high (a resistance of 15 Ohm and a capacitance of 100 pF) and the second (dashed black line) when the load is small (a resistance of 0.2 Ohm  and a capacitance of 1 pF). In both cases the traps are driven at a frequency of $\omega=2\pi\times20$ MHz. Figure \ref{antratio} shows the ratio of number of antenna turns to winding pitch as a function of the diameter of the coil required to impedance match the load to a 50 Ohm source. Figure \ref{antpitch} then illustrates how an existing antenna coil can be physically stretched (increasing the winding pitch of the coil) to impedance match to a source load. In figure \ref{antpitch} the number of turns is kept constant at 3 and the diameter is kept constant at 3 cm.

Figure \ref{antpitch} shows that in order to impedance match a source to a high impedance trap load (solid line) the antenna coil must be stretched compared with that required to match a low impedance trap load (dashed line).

\begin{figure}
\centering \includegraphics[width=1\columnwidth]{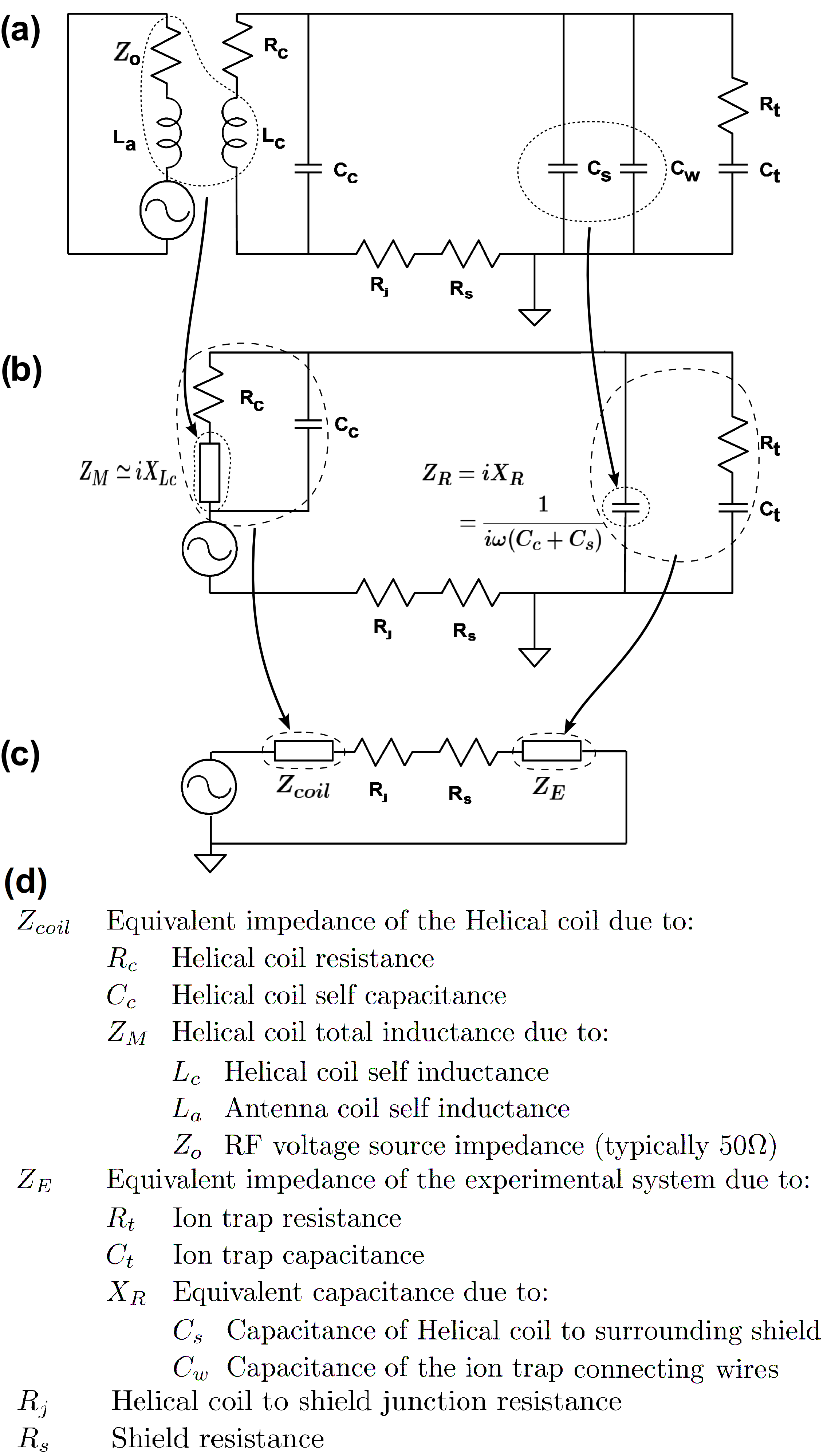}
\caption{Diagram showing the electrical equivalent of the overall resonant circuit. Part (a) shows the lumped element model electrical equivalent. Part (b) shows the simplified circuit. Part (c) shows the set of serial impedances the resonator can be represented as.}
\label{electrical_resonator}
\end{figure}

\subsection{Description of resonant frequency and $Q$ factor using an LCR circuit model}\label{sec:theory}
In order to calculate the $Q$ factor and resonant frequency $\omega_{0}$, the resonator is modelled as a lumped element circuit, shown in figure \ref{electrical_resonator}(a), which is simplified via figure \ref{electrical_resonator}(b) to figure \ref{electrical_resonator}(c) by creating Thévenin equivalent impedances, where each component is defined in the table in figure \ref{electrical_resonator}(d).

The coil impedance will depend on the mutual coupling, $Z_{M}$, which can be written as

\begin{equation}\label{Zm}
Z_{M}=iX_{Lc}+\frac{\omega^{2}M^{2}}{i\omega L_{a}+Z_{0}}
\end{equation}

\noindent
by using the same method employed to arrive at equation \ref{zin2}, where $X_{Lc}=L_{c}\omega$ is the reactance due to the inductance of the main coil. However, with the typical values required by ion trapping of RF drive frequency between $\omega\approx2\pi\times10$ MHz and $2\pi\times50$ MHz, $L_{c}\approx L_{a}\approx1$ mH and $Z_{0}=50$ Ohm it can be shown that $\left|Z_{M}\right|\approx iX_{Lc}$. Thus, we can express:

\begin{equation}\label{Zcoil}
Z_{coil}=\left(\frac{1}{(iX_{Lc}+R_{c})}+\frac{1}{iX_{Cc}}\right)^{-1}.
\end{equation}

\noindent
Summing the trap capacitance and resistance in parallel with the wire capacitance and shield capacitance we can write the $Z_{E}$ impedance as

\begin{equation}\label{Zload}
Z_{E}=\left(\frac{1}{(iX_{Ct}+R_{t})}+\frac{1}{iX_{Cw}}+\frac{1}{iX_{Cs}}\right)^{-1},
\end{equation}

\noindent
where $X_{C_{t}}=\frac{1}{C_{t}\omega}$ is the reactance of the trap capacitance and $X_{C_{w}}$ and $X_{C_{s}}$ are the reactance due to $C_{w}$ and $C_{s}$ respectively.

The total impedance of the resonator, $Z_{tot}$, can then be expressed as

\begin{equation}\label{Ztot}
Z_{tot}=Z_{coil}+Z_{E}+R_{s}+R_{j}.
\end{equation}

We can express both $Z_{coil}=R_{coil}+iX_{coil}$ and $Z_{E}=R_{E}+iX_{E}$, where $R_{coil}$ and $R_{E}$ are the equivalent series resistance of the coil and experimental system respectively and $X_{coil}$ and $X_{E}$ are the equivalent series reactance for the coil and experimental system respectively. At resonance $Z_{tot}$ will be purely resistive when
\begin{equation} 
X_{coil}+X_{E}=0\label{resonance_condition}.
\end{equation} 
From the calculated Thévenin equivalent impedances from figure \ref{electrical_resonator} it can be shown that at resonance,

\begin{equation}\label{resonance_condition_expanded}
\frac{i\omega_{0}L_{C}}{1-L_{C}C_{C}\omega_{0}^{2}}+\frac{1}{i\omega_{0}(C_{s}+C_{t}+C_{w})}=0.
\end{equation}

This equation can be used to calculate the resonant frequency:

\begin{equation}\label{resonance_frequency}
\omega_{0}=\frac{1}{\sqrt{(C_{s}+C_{t}+C_{w}+C_{C})L_{C}}}.
\end{equation}

The $Q$ factor is defined as $Q\equiv\omega_{0}\frac{\textrm{Energy stored}}{\textrm{Power dissipated}}$. The energy stored in the resonator will oscillate between the inductance of the coil and the combined capacitances in the circuit. The total energy stored will be equal to the peak energy stored in the inductor $E_{Lc}=I_{peak}^{2}L_{c}/2=I_{rms}^{2}L_{c}$ and the power dissipated in the system is $P_{d}=I_{rms}^{2}R_{_{ESR}}$ where $R_{_{ESR}}$ is the equivalent series resistance of the circuit and is given by the real part of equation \ref{Ztot}. It can then be shown that the $Q$ factor of a resonator is given by the following:

\begin{equation}\label{qualityfactor}
Q=\frac{X_{L_{c}}}{R_{_{ESR}}}.
\end{equation}

From equation \ref{Ztot}, $R_{_{ESR}}$ can be derived as

\begin{equation}\label{real_part}
R_{_{ESR}}=\frac{R_{c}X_{Cc}^{2}}{R_{c}^{2}+(X_{Cc}+X_{Lc})^{2}}+\frac{R_{t}X_{R}^{2}}{R_{t}^{2}+(X_{R}+X_{T})^{2}}+R_{s}+R_{j}.
\end{equation}

For helical coils with a low self capacitance ($X_{Lc}\ll X_{Cc})$ and low resistance ($R_{c}\ll X_{Cc})$, and for an ion trap with low resistance such that $R_{t}\ll(X_{Ct}+X_{Cw}+X_{Cs})$, this can be approximated as

\begin{equation}\label{real_part_simple}
R_{_{ESR}}\simeq R_{j}+R_{c}+R_{s}+R_{t}\alpha^{2},
\end{equation}

\noindent
where $\alpha=\frac{a}{a+1}$, $a=\frac{X_{R}}{X_{T}}=\frac{C_{t}}{C_{s}+C_{w}}$ is the ratio of the trap capacitance to the combined capacitance due to the connecting wires and coil shield. For $a\rightarrow0$ the capacitance of the shield and wires shunt the RF current and there is negligible contribution from the resistance of the ion trap. For $a\rightarrow\infty$ the trap capacitance dominates giving a maximum contribution of the ion traps resistance.

It should be noted that, as high Q factors are obtained when maximising the coil inductance and minimising the system capacitance, the effect of wire and feed-through inductance will be negligible compared to a typical coil inductance and, as such, has not been included in the model. This is not true of stray capacitances which can be on the same order of magnitude as the ion trap capacitance and can be treated as part of the wire capacitance $C_w$.

\subsection{Resonator Capacitance, Inductance and Resistance}

\begin{figure}
\centering \includegraphics[width=1\columnwidth]{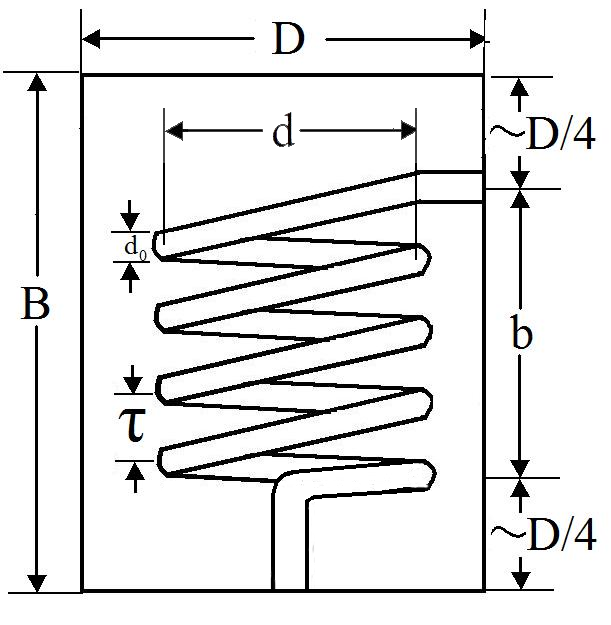}
\caption{Outline design of a resonator showing the shield diameter $D$, shield height $h$, coil diameter $d$, coil height $b$, winding pitch $\tau$ and the coil wire diameter $d_{0}$}
\label{resonator}
\end{figure}

In order to construct a resonator to operate at a desired frequency the capacitance of the wires and ion trap $C_{\Sigma}=C_{w}+C_{t}$, which depend on the configuration of the ion trap and experimental setup, can be measured at the vacuum systems feed-through with a capacitance meter. Once these are known we must construct a shielded coil with the necessary capacitance $C_{c}$ and $C_{s}$ and inductance $L_{C}$. All equations in this section assume the use of SI units unless otherwise stated. The self capacitance of the coil in units of farads, given empirically by Medhurst, is \cite{Medhurst}

\begin{equation}\label{med}
C_{C}\simeq(Hd)\times10^{-12}\textrm{ farads},
\end{equation}

\noindent
where $d$ is the diameter of the coil and $H$ is given empirically by, $H=11.26\frac{b}{d}+8+\frac{27}{\sqrt{\frac{b}{d}}}$ farads/metre.

The capacitance present between the coil wire and the outer shield in units of farads is given empirically by \cite{Macalpine}

\begin{equation}\label{Cs}
C_{s}\approx bK_{Cs}\left(d,D\right)\textrm{ farads},
\end{equation}

\noindent
where $K_{Cs}\left(d,D\right)=39.37\frac{0.75}{\log\left(\frac{D}{d}\right)}\times10^{-12}$ farads/metre, $d$ is the diameter of the coil, $D$ the inner shield diameter and $b$ is the height of the coil.

The inductance of a coil inside a shield in units of henrys, for $b/d\geq1$, is given empirically by \cite{Macalpine}

\begin{equation}\label{Lc}
L_{C}\approx bK_{Lc}\left(d,D,\tau\right)\textrm{ henrys},
\end{equation}

\noindent
where $K_{Lc}\left(d,D,\tau\right)=39.37\frac{0.025d^{2}(1-(\frac{d}{D})^{2})}{\tau{}^{2}}\times10^{-6}$ henrys/metre and $\tau$ is the winding pitch of the coil.

We can approximate the required height of a coil, $b$, for a resonator to operate at frequency $\omega_{0}$, for chosen parameters of $d$, $D$ and $\tau$ and for a set of measured capacitances $C_{\Sigma}=C_{w}+C_{t}$. The coil's self capacitance, given by equation \ref{med}, significantly complicates the solution for coil height, $b$, however a simpler solution can be found by approximating the self capacitance as a linear equation.
Examining equation \ref{med} it can be seen that the maximum of the non linear term $\frac{0.27}{\sqrt{b/d}}$ occurs when $b/d<1$. As we require that $b/d\geq1$ a simple linear approximation for the coil self capacitance can be found by setting the $\sqrt{b/d}$ term to 1.
This gives an overestimate of the self capacitance but allows for an approximate solution for the coil height which will give a resonator with a resonant frequency typically within $2\%$ of the desired frequency.
The linear approximation to the coil self capacitance is, $C_{C}\simeq K_{cb}b+K_{cd}$, where $K_{cb}=11.26\times10^{-12}$ farads/metre and $K_{cd}=35d\times10^{-12}$ farads. Substituting this approximation with equations \ref{Lc} and \ref{Cs} into equation \ref{resonance_frequency} and rearranging for $b$ in units of metres, we obtain:


\begin{eqnarray}\nonumber\label{eq:coilheight}
b&\simeq&\frac{C_{\Sigma}+K_{cd}}{K_{Cs}+K_{cb}}\left(\sqrt{\frac{K_{Cs}+K_{cb}}{(C_{\Sigma}+K_{cd})^2 K_{Lc} \omega_{0}^{2}}+\frac{1}{4}}\right.\\
&&\left.-\frac{1}{2}\right)\textrm{ metres}.
\end{eqnarray}

In order to estimate the resistance of the resonator one must consider the path along which the current flows and how it flows along this path. The current will flow along the surface on its path with a depth, $\delta$, given by the skin depth of the coil material (in this case copper) at the resonant frequency of the resonator. However, additionally, the current in the shield must form a spiral which acts to form an equal and opposite magnetic field as that produced by the coil (Lenz's law). The distance around the shield the current will travel, $l_{s}$, can be calculated by equating the magnetic field, created by the coil at the shields surface:

\begin{equation}\label{coilB}
B_{field}=\frac{\mu Il_c}{4\pi(D-d)^{2}},
\end{equation}

\noindent
to that created by the shield $B_{shield}$:

\begin{equation}\label{shieldB}
B_{shield}=\frac{\mu N_{s}I}{b},
\end{equation}

\noindent
where $l_{c}$ is the unwound length of the coil. 

This can then be solved to find the number of turns the current undergoes in the shield, $N_{s}$:

\begin{equation}\label{shieldturns}
N_{s}=\frac{bl_c}{4\pi(D-d)^{2}}.
\end{equation}

The distance the current will travel from the bottom of the shield to the top of the shield can then be calculated as

\begin{equation}
l_{s}=N_{s}\sqrt{(\pi^{2}D^{2})+\left(\frac{b}{N_{s}}\right)^{2}}.
\end{equation}

The resistance of the coil and the shield can now be calculated using the relationship between resistance, $R$, and resistivity, $\rho$:

\begin{equation}\label{resistivity}
R=\frac{\rho l_{s}}{A},
\end{equation}

\noindent
where $l$ is the length along which the current travels and $A$ the area through which the current travels. We can now describe the resistance of the coil and shield as

\begin{equation}\label{rcoil}
R_{c}=\frac{\rho l_{c}}{d_{0}\pi\delta},
\end{equation}

\begin{equation}\label{rshield}
R_{s}=\left(\frac{N_{s}\rho l_{s}}{b\delta}\right),
\end{equation}

\noindent
where $d_{0}$ is the diameter of the coil wire.

We must also take into account additional resistances acquired by attaching the coil to the shield. The coils in this paper are attached to the shield by soldering them onto a BNC bulkhead located at a distance $\frac{D}{4}$ from the top of the shield as indicated in \cite{Macalpine} and figure \ref{solderjoint}. The solder joint created by this method will provide an additional resistance. The resistance of the connection at an angular frequency, $\omega_{n}$ is given by

\begin{equation}\label{Rn}
R_{n}=\frac{\rho l}{\pi d_{j}\delta_{n}},
\end{equation}

\noindent
where $\rho$ is the resistivity of the material, $l$ is the length through which the current flows, $\delta_{n}$ is the skin depth at an angular frequency $\omega_{n}$ and $d_j$ is the diameter of the solder joint. Due to the effects of skin depth at high frequencies a simple DC resistance measurement of the joint connecting the coil to the shield is not useful. Instead an AC resistance measurement must be made of the joint. Equipment exists (for example ISO-TECH: LCR819) which can perform resistance measurements at frequencies of approximately 100 kHz and this measurement can then be used to infer the resistance of the joint at a higher frequency, in this case the resonant frequency of the resonator. The measurement frequency needs to be chosen so that the skin depth is smaller then the radius of the joint. By defining $\gamma=(\rho l)/(\pi d_{j})$ equation \ref{Rn} can be re-written as

\begin{equation}\label{k}
R_{n}=\frac{\gamma}{\delta_{n}}.
\end{equation}

By using the frequency independent parameter, $\gamma$, it is possible to show for two different angular frequencies, $\omega_{1}$ and $\omega_{2}$,

\begin{equation}\label{r}
R_{1}\delta_{1}=R_{2}\delta_{2},
\end{equation}

\noindent
and rearranging using $\delta_{n}=2\sqrt{(\rho)/(\omega_{n}\mu_{0})}$ gives the resistance at a resonant angular frequency $\omega_{0}$ in terms of the resistance measured at an angular frequency $\omega_{1}$:

\begin{equation}\label{r2}
R_{0}=R_{1}\sqrt{\frac{\omega_{0}}{\omega_{1}}}.
\end{equation}

This derivation is only valid in a frequency regime where the solder joint is larger than the skin depth and when the resistivity, $\rho$, of the material is constant over the two frequencies used. This is the case for the resonators made in this paper as the skin depth is of the order of 10 $\upmu\mathrm{m}$ and the solder joint size is on the order of a few millimetres.

Equation \ref{r2} shows that by taking a resistance measurement at one frequency it is possible to calculate the resistance at another frequency. This method was used to calculate the resistance of the connection made between the main coil and the shield at the resonators resonant frequency.

\section{Resonator design guide and analysis}
\subsection{Design guide}
This section will provide a design guide which will enable a helical resonator to be constructed with a $Q$ factor close to that of the highest Q possible for a given set of parameters consisting of the desired resonant frequency, $\omega_{0}$, ion trap capacitance, $C_{t}$, and resistance, $R_{t}$, wire capacitance, $C_{w}$ and the size constraints for the resonator. The helical resonator may require different construction techniques depending on priorities set by these constraints, however there are some design issues universal to any resonator that must be considered. 

When designing a resonator it is important that certain constraints are met for the resonant frequency and $Q$ factor to be predicted by the theory. The resonant frequency depends on the inductance, which is predicted by equation \ref{Lc}. For this equation to be valid the coil height should be greater than the coil diameter, $b\geq d$, and the coil diameter should be greater than the wire diameter, $d\geq d_{0}$. 

We can see from equation \ref{Lc} that both the coil diameter and the winding pitch have a strong ($d^2$ and $1/\tau^2$) effect on the inductance. It is important to ensure the coil is made with precision for the winding pitch and the diameter of the coil to be constant along its length. This dependence also requires that the coil is not susceptible to mechanical vibrations. The strong effect the winding pitch, $\tau$, has on the inductance will result in vibrations of the coil causing the inductance and hence resonant frequency of the resonator to become time dependent. In order to minimise vibrational effects the coil should be constructed to be rigid and should be firmly mounted inside the shield. Finally, the coil must be mounted centrally inside an outer shield of height $B\geq b+D/2$ (we would typically recommend $B=b+D/2$), where b is the coil height and D is the shield diameter, in order to keep the coil fringe effects from reducing the coil's inductance and increasing the shield capacitance \cite{Zverev}.

In order to achieve high $Q$ factors the resonator must be built to minimise the resistance of the shield, the coil and solder joints. The coil and shield should be made from a highly conductive material (such as copper) which is thicker than the skin depth at the desired operating frequency. Any solder junctions should be made with a clean oxide free surface before soldering with both parts of the joint reaching a sufficient temperature to ensure good solder flow between them.

\begin{figure}
\centering \includegraphics[width=1\columnwidth]{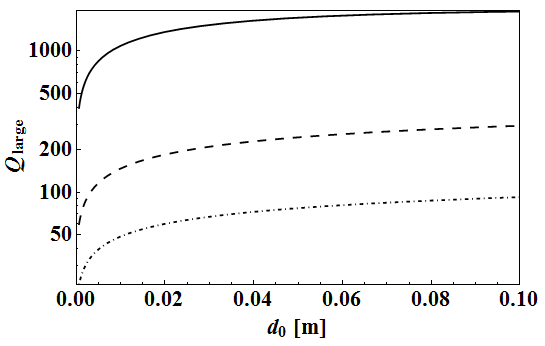}
\caption{$Q$ factor for values of $d$ and $d/D$ that maximise the $Q$ factor for varying wire diameter, $d_{0}$, for traps D (solid), H (dashed) and I (dot-dashed) from table \ref{spec_traps} at $\omega_0=2\pi\times10$ MHz}
\label{Qvsd0}
\end{figure}

\begin{figure}
\centering \includegraphics[width=1\columnwidth]{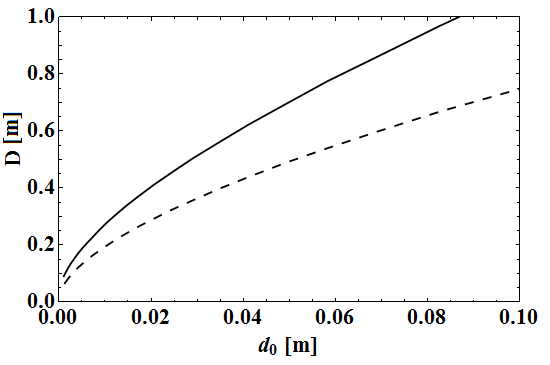}
\caption{$D$ vs $d_{0}$ for the $D$ value that achieves a $Q$ factor of $Q_{large}$ (solid line) and the minimum $D$ value that achieves a $Q$ factor of $Q_{90\%}$ (dashed line) for trap D from table \ref{spec_traps} at $\omega_0=2\pi\times10$ MHz}
\label{Dvsd0}
\end{figure}

A low resistance for the helical coil can be obtained by ensuring the use of a large diameter wire, $d_{0}$. The effect of the wire diameter on the $Q$ factor can be seen by plotting the `largest' $Q$ factor vs wire diameter $d_{0}$. We can define the `largest' $Q$ factor available, $Q_{large}$, for a given set of parameters, $\omega_{0}$, $C_{t}$, $C_{w}$, $R_{t}$ and $d_{0}$, where the coil diameter, $d$, and shield diameter, $D$, are chosen to maximise the $Q$ factor. 
Figure \ref{Qvsd0} shows a plot of $Q_{large}$ vs $d_0$ for three traps of $C_t=5$ pF and $R_t=5$ Ohm, $C_t=20$ pF and $R_t=15$ Ohm, $C_t=50$ pF and $R_t=15$ Ohm all for a resonant frequency of $\omega_0=2\pi\times10$ MHz. There is an asymptotic increase to higher $Q$ factors for large values of $d_0$. Even for high resistance, high capacitance traps modest $Q$ factors of $\approx$100 can be achieved, however this requires the coil to be formed from a thick rod. The upper limit to $d_{0}$ will result from an intersection of the limits that the coil height must be larger than the coil diameter, $b/d\geq1$, and the coil diameter must be larger than the wire diameter, $d>d_{0}$. Increasing $d_{0}$ will increase the $Q$ factor but will also increase the size of the resonator. It can be seen in figure \ref{Dvsd0} how the shield diameter required for $Q_{large}$ (solid line) rapidly increases with $d_{0}$. However, for a $Q$ factor 90$\%$ of $Q_{large}$, $Q_{90\%}$, a smaller shield diameter can be used. The minimum $D$ for $Q_{90\%}$ is shown in figure \ref{Dvsd0} (dashed line). It is clear that making a resonator with a $Q$ factor of $Q_{90\%}$ can reduce the size required for the resonator.

\begin{figure*}
\centering \scalebox{1}[1]{\includegraphics[width=2\columnwidth]{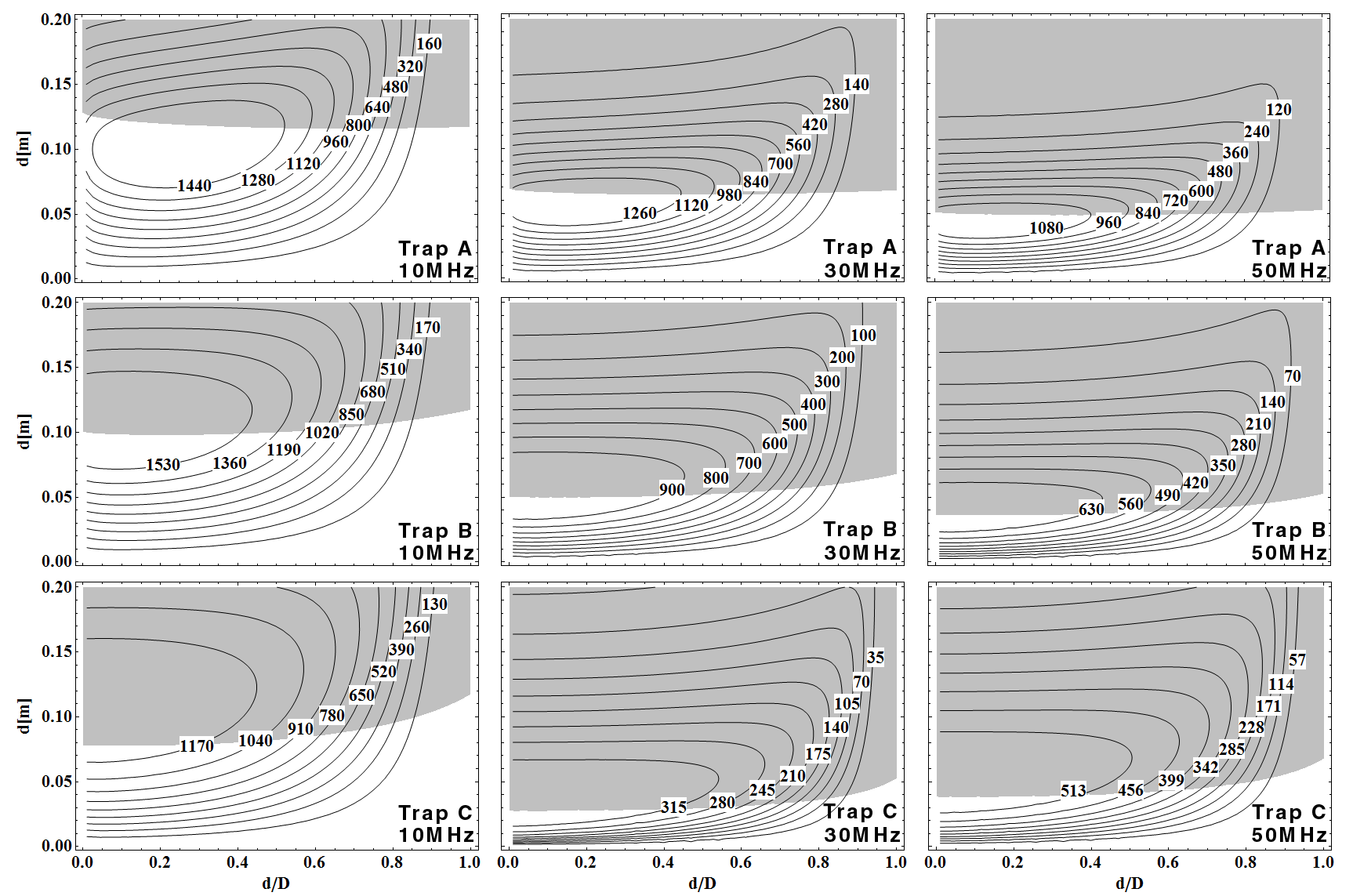}}
\caption{Contour plots for traps A, B and C (specifications in table \ref{spec_traps}) shown for operation frequencies of $\omega_0=2\pi\times10$ MHz, $\omega_0=2\pi\times30$ MHz and $\omega_0=2\pi\times50$ MHz. The grey areas indicate where $b/d<1$ therefore invalidating the theory.}
\label{trapABC}
\end{figure*}

\begin{figure*}
\centering \scalebox{1}[1]{\includegraphics[width=2\columnwidth]{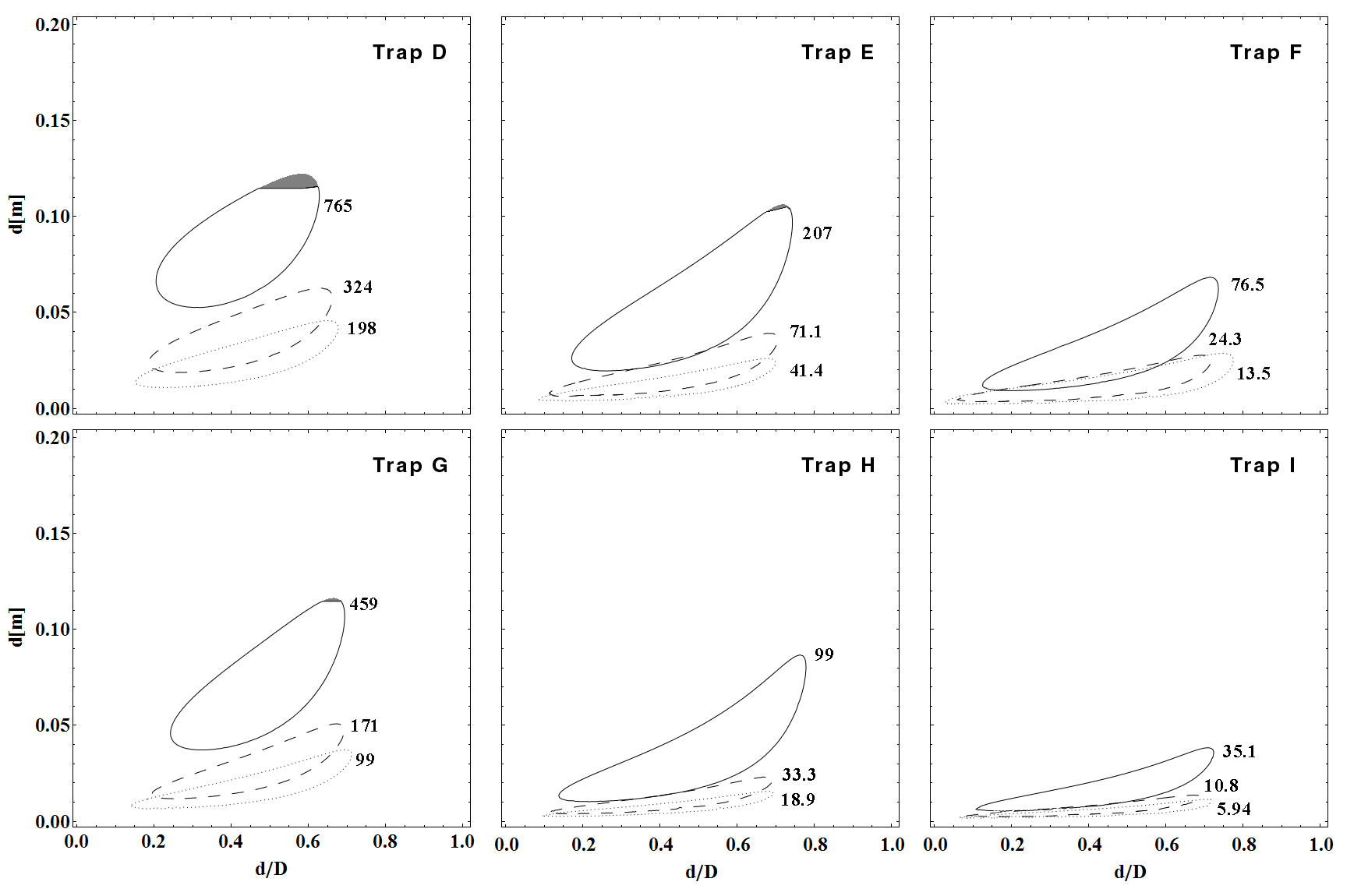}}
\caption{Contour plots showing $Q_{90\%}$ for each set of paramters coresponding to traps D to I from table \ref{spec_traps} for operating frequencies of $\omega_0=2\pi\times10$ MHz (solid lines), $\omega_0=2\pi\times30$ MHz (dashed lines) and $\omega_0=2\pi\times50$ MHz (dotted lines). The grey areas indicate where $b/d<1$ therefore invalidating the theory. The value of $Q_{90\%}$ is indicated next to the contour line.}
\label{trapDEF}
\end{figure*}

When designing a helical resonator for a specific experiment it is useful to examine contour plots of the $Q$ factor as a function of coil diameter, $d$, and the ratio of the coil diameter to the shield diameter, $d/D$, as shown in figure \ref{trapABC} and \ref{trapDEF}. Using these plots it is possible to choose values of $d$ and $d/D$ that will optimise the $Q$ factor for a set of parameters. These plots can be obtained using the parameters for $\omega_{0}$, $C_{t}$, $C_{w}$, $R_{t}$ for a given experiment and choosing values of $d_{0}$, $\tau$ and measuring or estimating $R_{j}$. The $Q$ factor $Q(d,d/D)$ can be obtained by calculating: 
\begin{description}
	\item[$b(d,d/D)$]- The coil height - by using equation \ref{eq:coilheight}, $K_{Cs}$ \& $K_{Lc}$ from equation \ref{Cs} \& \ref{Lc} and $K_{cb}$ \& $K_{cd}$ from the approximation for the coil capacitance.
	\item[$C_{s}(d,d/D)$]- The coil to shield capacitance -  by substituting $b(d,d/D)$ into equation \ref{Cs}
	\item[$C_{c}(d,d/D)$]- The coil self capacitance - by substituting $b(d,d/D)$ into equation \ref{Cs}
	\item[$R_{s}(d,d/D)$]- The shield resistance - by using equation \ref{rshield}
	\item[$R_{c}(d,d/D)$]- The coil resistance - by using equation \ref{rcoil}
	\item[$R_{ESR}(d,d/D)$]- The total resistance - by using equation \ref{real_part_simple} 
	\item[$L_{c}(d,d/D)$]- The coil inductance - by substituting $b(d,d/D)$ into equation \ref{Lc}
	\item[$Q(d,d/D)$]- The $Q$ factor - by substituting $L_{c}(d,d/D)$ $\&$ $R_{ESR}(d,d/D)$ into equation \ref{qualityfactor}
\end{description}

Figures \ref{trapABC} and \ref{trapDEF} show contour plot for parameters corresponding to traps from table \ref{spec_traps} for three resonant frequencies with $d_{0}=5$ mm and $\tau=2d_{0}$. While larger values of $d_{0}$ would result in larger $Q$ factors, and a larger $Q_{large}$, the values for $d_0$ and $\tau$ have been chosen as they are typical for a hand wound coil, which will be discussed in detail later.
Both plots feature a grey shaded region indicating where the condition that $b/d\geq1$ (from equation \ref{Lc}) is not valid. Within this region the coil inductance will deviate from that predicted by equation \ref{Lc} resulting in a deviation from the predicted $\omega_{0}$ and $Q$ factor. An expansion to this theory for short helical coils or wire loops could be implemented by utilising an appropriate equation for the inductance.

  The experimental size restrictions will strongly dictate the achievable $Q$ factor of the resonator. The
values of $d$ and $d/D$ must be chosen to ensure that the shield diameter $D$ and the shield height $B$ will be within these size constraints, otherwise different values for $d$, $d/D$ or $d_{0}$ will need to be chosen. Figure \ref{trapABC} for traps A to C show contour lines at a range of $Q$ factors (as labelled) enabling the values of $d$ and $D$ to be chosen to maximise the $Q$ factor for a given size constraint. Figure \ref{trapDEF} for traps D to I shows contour lines of $Q_{90\%}$ for each set of parameters. The values of $d$ and $D$ can be chosen within the $Q_{90\%}$ contour line to optimise the $Q$ factor (for a set of parameters $\omega_{0}$, $C_{t}$, $C_{w}$, $R_{t}$, $R_{j}$, $\tau$ and $d_{0}$), while enabling a choice of $d$ and $D$ that minimises the size of the resonator.

For traps A, B and C figure \ref{trapABC} shows that higher $Q$ factors are achieved when the coil diameter to shield diameter ratio, $d/D$, is close to 0. This corresponds to a large separation between the coil and the shield and hence a small shield capacitance. At $d/D=0$, the shield diameter, $D$, would be infinite and the shield capacitance would be zero, indicating the resonator is dominated by the trap capacitance. However, we can see in figure \ref{trapDEF} that $Q$ factors of $Q_{90\%}$ can be achieved at values of $d/D$ of order 0.5 with larger trap capacitance enabling larger ratios and hence a smaller shield diameter $D$ and higher shield capacitance.

While figures \ref{Qvsd0} and \ref{Dvsd0} show that at large values of $d_{0}$ high $Q$ factors may be achieved at a given size constraint, the construction of such resonators needs to be taken into account. In order to construct a resonator without specialist machinery, a wire diameter of approximately 5 mm is recommended. This wire size is sufficiently rigid not to be susceptible to mechanical vibrations, while being flexible enough when heated to be wound by hand into a coil. This can be achieved by winding the wire around a tube with notches cut into it to help align the wire to a constant winding pitch. The size of the resonator can be reduced by using a small winding pitch, however, a minimum winding pitch of $\tau=2d_{0}$ is recommended when winding by hand in order to keep the error in the winding pitch small. A higher $Q$ factor could be achieved by using $d_{0}=10$ mm, however, this would be hard to wind by hand, which can result in large errors in the winding pitch. 

Coils can be constructed using large diameter wire but may require the use of specialist machinery. It should be noted that coils could be constructed from tubular material as current is only carried on the skin of the metal, however, this would affect the inductance of the coil. Similarly a rectangular cross-section wire could be used to form a coil or formed by cutting a tube into a coil. In order to predict $\omega_{0}$ and the $Q$ factor reliably equation \ref{Lc} may need to be replaced with an expression for the inductance suitable for the desired geometry.  \linebreak[4] 
 
\begin{description}
\item[]We can summarise:
\begin{itemize}
  \item A highly conducting material should be used to construct the resonator (for example copper).
	\item The coil wire should be made reasonably thick to provide mechanical stability and reduce coil resistance. If winding the coil by hand a wire on order of $d_{0}\approx5$ mm is suggested.
	\item The winding pitch should be as small and uniform as possible. If winding the coil by hand a minimum of $\tau\sim2d_{0}$ is recommended.
	\item A contour plot of $Q(d,d/D)$ can be used to determine appropriate parameters for d and D within size constraints.
	\item The coil height is calculated from equation \ref{eq:coilheight}.
	\item The coil height must be greater than the coil diameter for equation \ref{Lc} to be valid.
	\item The shield height $B$ should be $b+D/2$.
	\item The coil and shield should be connected as close to the vacuum system as possible.
	\item Any solder joints made should be of low resistance.
\end{itemize}
 \end{description} 

\begin{table}[ht]
\centering \caption{Specifications of traps used for figures \ref{trapABC} and \ref{trapDEF}}
		\begin{tabular}{|c|c|c|}
					 \hline
		       Trap & Resistance & Capacitance \\
		        & Ohm & pF \\
           \hline
           A & 0.1 & 5 \\  
           B & 0.1 & 20 \\
           C & 0.1 & 50 \\
           D & 5 & 5 \\
           E & 5 & 20 \\ 
           F & 5 & 50 \\
           G & 15 & 5 \\
           H & 15 & 20 \\
           I & 15 & 50 \\
           \hline
        \end{tabular}\label{spec_traps}
\end{table}

\subsection{Case Study}
\subsubsection {Resonator construction and measurement}
We will now look at a how a resonator for a typical ion trap experiment can be built without the need for specialist equipment. We will then discuss how the resonant frequency and $Q$ factor can be measured, while ensuring impedance matching between the RF source and the resonator.

The coil can be wound by hand by using an annealed copper wire of diameter, $d_{0}\approx5$ mm. The copper can be annealed by heating with a blow torch in order to give increased flexibility. Once cooled the copper can be wound, which will work harden the copper, creating a rigid coil. To ensure all the turns are equally spaced, the copper should be wound around a tube of diameter, $d-d_{0}$, with notches cut into the tube to align the wire when winding. 
\begin{figure}
\centering \includegraphics[width=1\columnwidth]{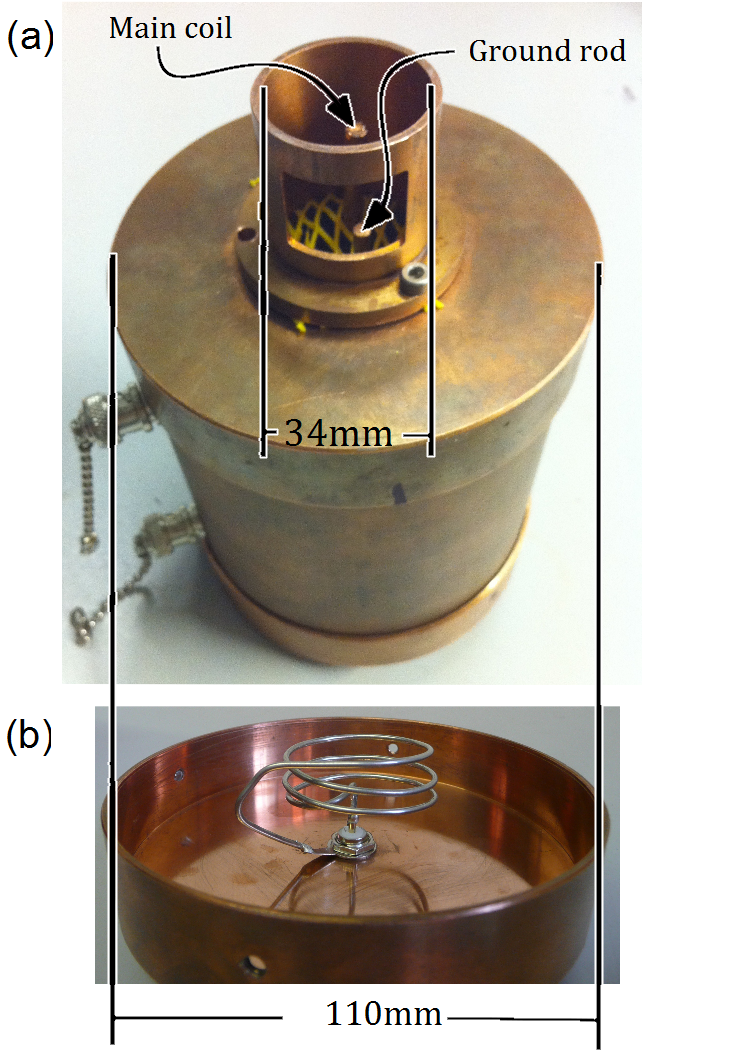}
\caption{(a) Picture showing the bottom end cap including the top hat used to connect the resonator to a vacuum feed-through (in this case Kurt J. Lesker: EFT 0523052). The main coil and ground rod can be seen exiting the resonator and are held in position by a plastic mesh. The window in the top hat provides access for connecting the main coil and ground rod to feed-through pins. (b) Picture showing the top end cap and antenna coil (shown in further detail in figure \ref{antenna}).}
\label{composite1}
\end{figure}

\begin{figure}
\centering \includegraphics[width=1\columnwidth]{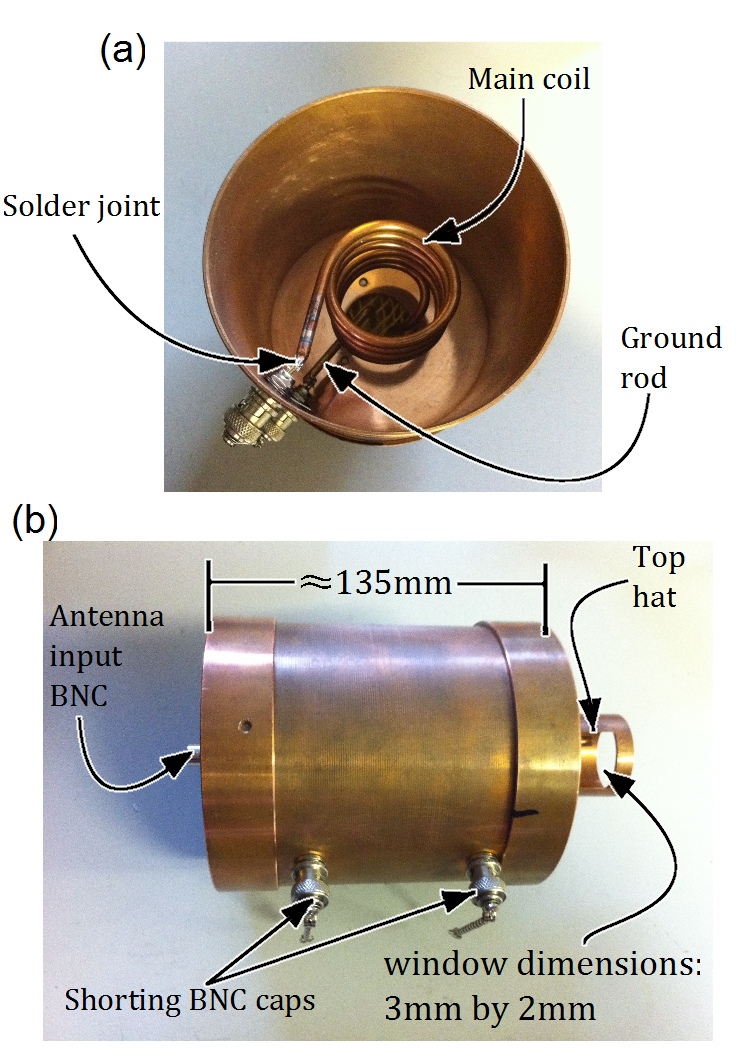}
\caption{(a) Picture showing a coil and location of the solder joint used to make an electrical connection to the shield. This is done with the use of a BNC bulkhead located at a distance $\frac{D}{4}$ from the top of the shield where $D$ is the diameter of the shield. The grounding rod is also shown. (b) Picture showing a fully constructed resonator. The top hat is shown on the bottom end cap and is designed to fit around a vacuum feed-through (in this case Kurt J. Lesker: EFT 0523052) which connects the main coil and grounding rod to the ion trap. A window is cut into the top hat to allow the connection between the feed-through and the main coil and grounding rod to be made. The top end-cap shows the BNC connection to the antenna coil, this is where the RF signal is applied to the resonator.}
\label{solderjoint}
\end{figure}

Once the coil is constructed it should be placed centrally inside the shield in order to minimise the coil to shield capacitance $C_{s}$. To ensure this, it must be clamped in place at the end of the coil before soldering the coil to a BNC bulkhead located in the shield as shown in figure \ref{solderjoint}. This clamping must be kept in position until the joint to the BNC is solid enough to support the coil on its own. This BNC bulkhead can be used to electrically connect the coil to the shield by connecting a BNC shorting cap. The ground rod is connected in the same way although to the lower BNC bulkhead shown in figure \ref{solderjoint}. The ground rod must exit the resonator (as shown in figure \ref{composite1}) without coming into contact with conducting material. The ground rod and main coil can be held in place with the use of a non-conducting mesh, as shown in figure \ref{composite1}, to reduce the mechanical stress applied to them from connection to a vacuum feed-through or other similar load.

\begin{figure}
\centering \includegraphics[width=1\columnwidth]{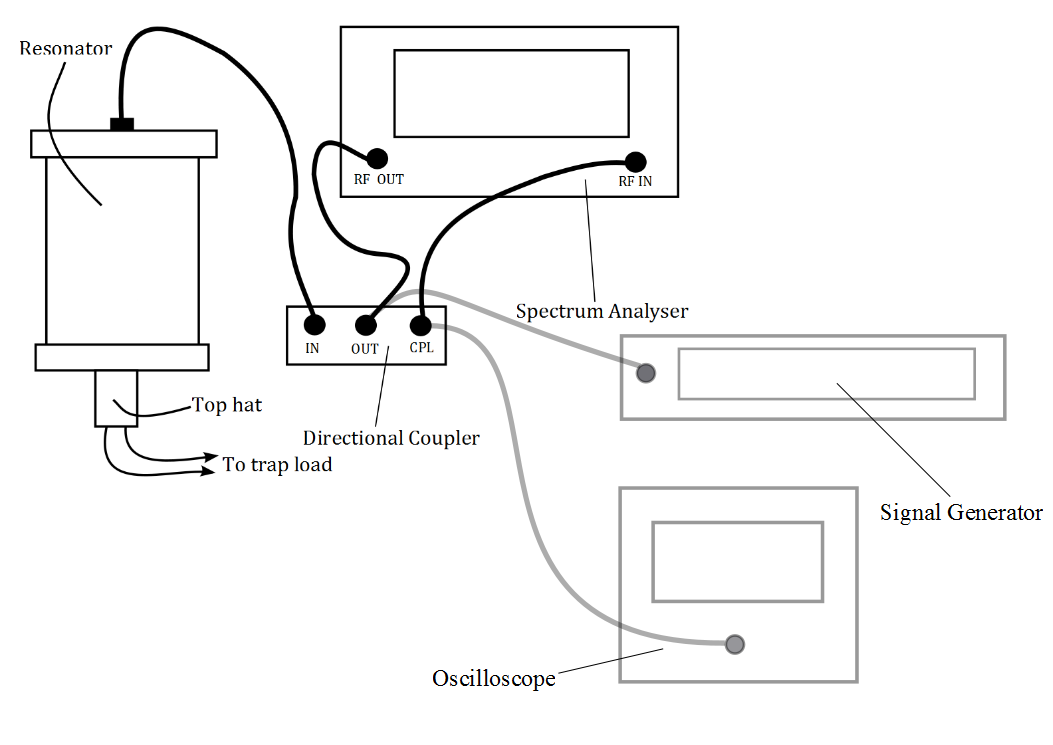}
\caption{Experimental set up required to measure the resonant frequency, coupling and $Q$ factor of a resonator. The resonator is connected to a spectrum analyser with a tracking generator via a directional coupler such that the reflected signal from the resonator is displayed on the spectrum analyser. Alternative equipment can be used and is shown in grey. This comprises of a signal generator and an oscilloscope.}
\label{setup}
\end{figure}

The antenna coil used to couple the radio frequency source to the resonator can be constructed out of $1.5\pm0.5$ mm wire and wound into $3$ turns with a winding pitch of $10\pm1$ mm and diameter of $33\pm1$ mm. This, however, should be varied in order to match the impedance of the resonator to the source, as described in figures \ref{antratio} and \ref{antpitch} and equation \ref{zin_antenna}. To measure the resonant frequency of the resonator, a directional coupler (Mini Circuits: ZDC-20-3) should be placed between the output source port and the RF input port of a spectrum analyser's tracking generator (Agilent: N9320B) as shown in figure \ref{setup}. Alternatively, the spectrum analyser can be replaced by an RF source and an oscilloscope, as shown in grey in figure \ref{setup}. The resonator should be connected to the directional coupler and RF source via the end-cap that hosts the antenna coil shown in figure \ref{antenna}. Using the experimental set up shown in figure \ref{setup} the resonant frequency results in a minimum in the spectral response of the reflected signal detected by the spectrum analyser. The pitch and diameter of the antenna coil should be altered until less than $5\%$ of the applied radio frequency signal is reflected back to the signal generator from the resonator when on resonance. The $Q$ factor of this resonance is simply measured by dividing the resonant frequency, $\omega_{0}$, by the full width of the resonance at $1/\sqrt{2}$ of the maximum voltage reflection, $\delta \omega_{0}$:

\begin{equation}\label{expQ}
Q=\frac{\omega_{0}}{\delta \omega_{0}}.
\end{equation}

Using the method described here it is possible to measure the resonant frequency and $Q$ factor of a resonator when the ion trap is unconnected, which corresponds to $R_{T}$ and $C_{T}$ being equal to infinity and zero respectively. The resonant frequency and $Q$ factor of a resonator with an ion trap applied across the output can then be measured by adding the required values of resistance and capacitance across the output of the resonator. The stray capacitance $C_{W}$ created between the wires used to connect the trap resistance and capacitance can be reduced by keeping these wires as short and as separated as possible.

\subsubsection {Experimental analysis of typical resonators}

\begin{figure}
\centering \includegraphics[width=1\columnwidth]{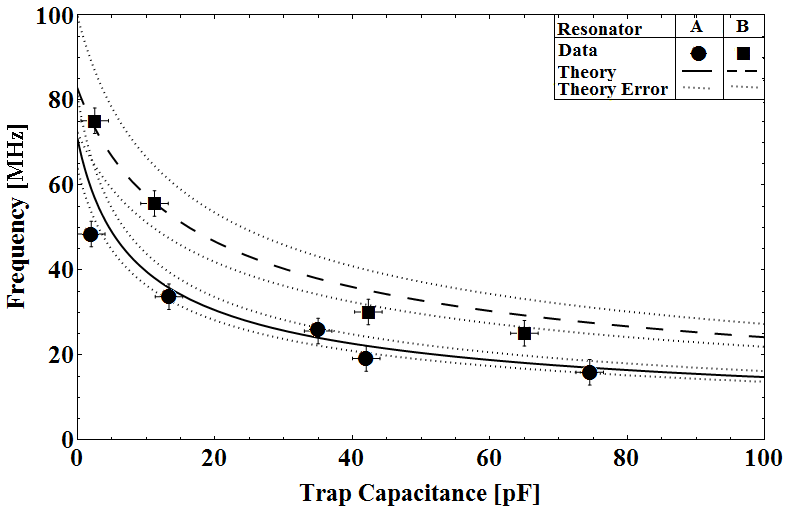}
\caption{The resonant frequencies of resonator A (circles) and resonator B (squares) are shown as a function of the trap capacitance they are attached to. The dashed curves represent the error on this calculation based on the design errors stated in table \ref{restable}. The resonant frequencies were measured for a resistance of $1$ Ohm, however, we note that they are actually independent of the resistance.}
\label{combinedF}
\end{figure}

\begin{figure*}
\centering \includegraphics[width=2\columnwidth]{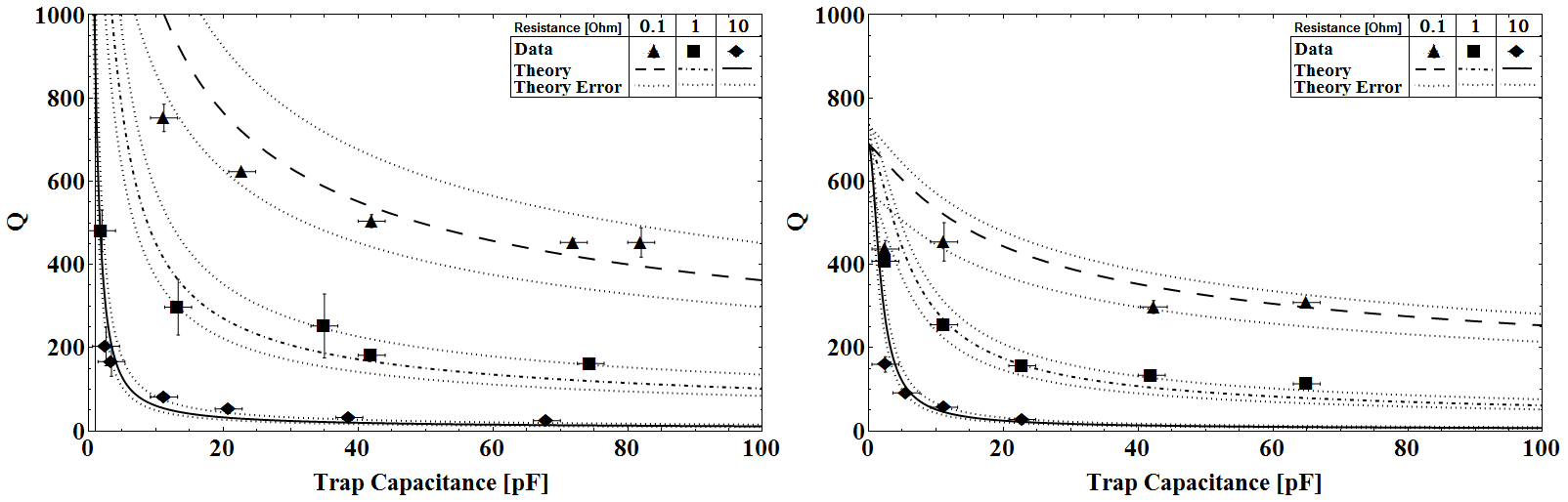}
\caption{The $Q$ factor of resonator A (left) and resonator B (right) are shown as a function of the trap capacitance. The dashed curves represent the error of the calculation based on the design errors stated in table \ref{restable}. The $Q$ factor was measured for different resistance loads shown by the triangles ($0.1$ Ohm), squares ($1$ Ohm) and diamonds ($10$ Ohm).}
\label{combinedQ}
\end{figure*}

Two resonators were constructed with a range of parameters as described in table \ref{restable}. The theoretical resonant frequencies of the resonators have been plotted in figure \ref{combinedF} as a function of the trap capacitance, $C_{t}$. The theoretical $Q$ factor for these resonators can be seen in figure \ref{combinedQ} plotted as a function of of the trap capacitance, $C_{t}$, applied to the resonator. All these are plotted for typical trap resistances, $R_{t}$, of $0.1$ Ohm, $1$ Ohm and $10$ Ohm, which is representative of the typical range over which the resistance of an ion trap can vary depending on what type of material and fabrication techniques are used.

\begin{table}\centering \caption{Specifications of the resonators. The $Q$ factors and frequencies quoted are without the addition of an ion trap load.}
	{\footnotesize }\begin{tabular}{|r|c|c|}
	\hline
	{\footnotesize Resonator } & {\footnotesize A } & {\footnotesize B }\tabularnewline
	\hline
	 &  & \tabularnewline
	{\footnotesize Shield diameter $D$ {[}mm{]} } & {\footnotesize 108$\pm$2 } & {\footnotesize 76$\pm$2 }\tabularnewline
	 &  & \tabularnewline
	{\footnotesize Shield height $h$ {[}mm{]} } & {\footnotesize 120$\pm$2 } & {\footnotesize 90$\pm$2 }\tabularnewline
	 &  & \tabularnewline
	{\footnotesize Coil diameter $d$ {[}mm{]} } & {\footnotesize 42$\pm$2 } & {\footnotesize 46$\pm$2 }\tabularnewline
	 &  & \tabularnewline
	{\footnotesize Coil wire diameter $d_{0}$ {[}mm{]} } & {\footnotesize 5.0$\pm$0.1 } & {\footnotesize 5.0$\pm$0.1 		}\tabularnewline
	 &  & \tabularnewline
	{\footnotesize Winding pitch $\tau$ {[}mm{]} } & {\footnotesize 9$\pm$1 } & {\footnotesize 15$\pm$1 }\tabularnewline
	 &  & \tabularnewline
	{\footnotesize Number of turns $N$ } & {\footnotesize 6.75$\pm$0.25 } & {\footnotesize 4.5$\pm$0.25 }\tabularnewline
	 &  & \tabularnewline
	{\footnotesize $d/D$ ratio } & {\footnotesize 0.4$\pm$0.2 } & {\footnotesize 0.6$\pm$0.2 }\tabularnewline
	 &  & \tabularnewline
	\hline
 	&  & \tabularnewline
	{\footnotesize Predicted frequency {[}MHz{]} } & {\footnotesize $64_{-6}^{+8}$ } & {\footnotesize $78_{-7}^{+10}$ }\tabularnewline
	 &  & \tabularnewline
	{\footnotesize Measured frequency {[}MHz{]} } & {\footnotesize $67\pm0.5$} & {\footnotesize $83\pm0.5$}\tabularnewline
	 &  & \tabularnewline
	{\footnotesize Predicted $Q$ } & {\footnotesize $1970_{-374}^{+252}$ } & {\footnotesize $689_{-115}^{+46}$ }\tabularnewline
	 &  & \tabularnewline
	{\footnotesize Measured $Q$ } & {\footnotesize 2176$\pm200$ } & {\footnotesize 631$\pm60$ }\tabularnewline
 &  & \tabularnewline
\hline
\end{tabular}\label{restable}
\end{table}
 
Figures \ref{combinedF} and \ref{combinedQ} show that the experimental measurement of the $Q$ factor and resonant frequency over a wide range of trap loads is consistent with the theory described in this work. The $Q$ factor of resonator A can be seen to be higher than that of resonator B as it possesses specifications which are either optimum or nearer to optimum than resonator B (depending on the trap load applied). It can be seen that although the resonator is not optimum for the various ion trap impedances, the $Q$ factor may still be sufficient for many experiments. This shows that a new resonator does not necessarily have to be built if the trap is altered slightly.

\section{Experimental Measurement of $\kappa$}\label{sec:expkappa}
It was shown in section 2 that the voltage output of a resonator is given by $V_{rms}=\kappa\sqrt{PQ}$, where $P$ is the power of the signal applied to the resonator, $Q$ is the quality factor of the resonator and $\kappa=(L/C)^{\frac{1}{4}}$. Here we will experimentally measure the value of $\kappa$ in an ion trap experiment. The value of $\kappa$ is required in order to calculate the voltage applied to the ion trap electrodes used to create a trapping potential. A resonator, described in table \ref{specs}, was electrically connected to an ion trap and vacuum system with a capacitance and resistance measured to be $17\pm2$ pF and $\approx0.1$ Ohm respectively. The resonant frequency and $Q$ was then measured with this additional load to be $\omega_0=2\pi\times21.895\pm0.010$ MHz and $Q=477\pm28$ respectively.

\begin{table}
\centering \caption{Specification of resonator used for $\kappa$ measurements\label{specs} }

\begin{tabular}{|r|c|}
\hline
Resonator & C\tabularnewline
\hline
\hline
Shield diameter, $D$ {[}mm{]} & $76\pm1$ \tabularnewline
 & \tabularnewline
Shield height,  $B$ {[}mm{]} & $105\pm1$ \tabularnewline
 & \tabularnewline
Coil diameter, $d$ {[}mm{]} & $52\pm2$\tabularnewline
 & \tabularnewline
Coil wire diameter, $d_{0}$ {[}mm{]} &  $4.0\pm0.1$ \tabularnewline
 & \tabularnewline
Winding pitch, $\tau$ {[}mm{]} & $7\pm3$ \tabularnewline
 & \tabularnewline
number of turns, $N$ & $7\frac{3}{4}$\tabularnewline
\hline
\end{tabular}

\end{table}

\begin{figure}
\centering \includegraphics[width=1\columnwidth]{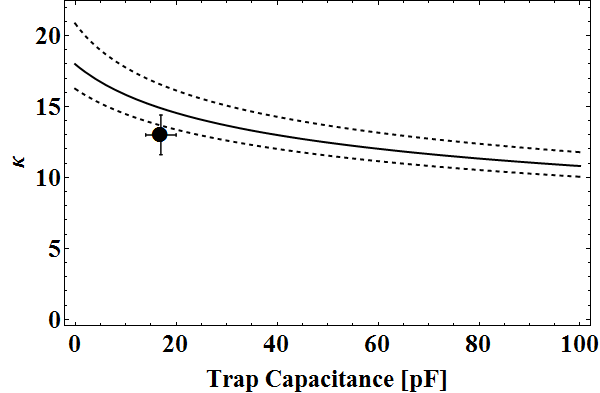}
\caption{The factor $\kappa$ from equation \ref{kappa} is plotted as a function of the the trap capacitance for the resonator described in section \ref{sec:expkappa}. The thick curve shows the value of $\kappa$ with the dashed curves showing the error on this calculation due to the uncertainty of the resonator specifications and its imperfections. The data point shown is for the resonator attached to a $17\pm3$ pF ion trap and vacuum system.}
\label{kappagraph}
\end{figure}

A single $^{174}Yb$ ion was trapped in the electric field created by the trap electrodes with the resonator supplied with $1.0\pm0.1$ W at its resonant frequency. The secular frequencies of the ion under these conditions were then measured and a boundary element method (BEM) model of the trapping field was used to determine the RF voltage required to create such a field. This voltage was found to be $400\pm20$ V which, when used in equation \ref{kappa} along with the power used to trap the ion, results in a $\kappa$ of $12.9\pm1.4$. This result is compared with the theoretical prediction of $\kappa$ (from equation \ref{kappa}) in figure \ref{kappagraph}. We note that the value of $\kappa$ depends on the impedance of the ion trap attached to the resonator.

\section{Conclusions}

We have carried out a detailed study of helical resonators for the use in applying high voltages at radio frequencies to ion traps. This has been done by modelling the resonator as a lumped element circuit along with a detailed discussion on the losses present in helical resonators connected to ion trap loads in order to arrive at an expression for the $Q$ factor and resonant frequency $\omega_{0}$. It has been shown how a resonator and load can be impedance matched to a frequency source by simply adjusting the physical parameters of an antenna coil which inductively couples the two. A general expression for the voltage output of the resonating system has been derived in terms of the systems $Q$ factor, input power, $P$, and a $\kappa$ factor which is a function of the systems capacitance and inductance. We have experimentally confirmed the value of this factor using a single trapped ion. The theory described in this paper has been confirmed by fabricating two resonators and measuring their $Q$ factor and resonant frequency, $\omega_{0}$, for a range of different trap loads ($C_{T}$ from 2 pF to 85 pF and $R_{T}$ from $0.1$ Ohm to $10$ Ohm).

A detailed design guide has been presented showing how a helical resonator can be designed which provides the highest $Q$ factor achievable for a desired resonant frequency within the constraints of a particular experiment. Producing a resonator with a optimised $Q$ factor allows the application of high voltages with optimised filtering. This will result in less noise injected into the system which could reduce anomalous heating of the trapped ions. The application of high voltages can give larger trap depths leading to longer trapping lifetimes and increased secular frequencies. This technology plays a significant role in many ion trapping experiments and should allow for progress in a variety of fields that require trapped ions or high radio frequency voltages.

\section*{ACKNOWLEDGEMENTS}
This work was supported by the UK Engineering and Physical Sciences Research Council (EP/E011136/1 and EP/G007276/1), the European Commission\textquoteright{}s Sixth Framework Marie Curie International Reintegration Programme (Grant No.MIRG-CT-2007-046432), the Nuffield Foundation, and the University of Sussex.


\end{document}